\documentclass[aps,prb,twocolumn,nobibnotes,nofootinbib,longbibliography,superscriptaddress]{revtex4-2}
\usepackage{mathrsfs}
\usepackage{amsmath,textcomp,gensymb}
\usepackage{amsfonts}
\usepackage{amssymb}
\usepackage{amsthm}
\usepackage{graphicx}
\usepackage{natbib}
\usepackage{color}
\usepackage[colorlinks=true,citecolor=blue,linkcolor=blue, urlcolor=cyan]{hyperref} 
\usepackage{bm}
\usepackage[caption=false]{subfig}
\usepackage{verbatim}

\usepackage{xcolor,cancel}

\usepackage{empheq}
\definecolor{myblue}{rgb}{.8, .8, 1} 
    
\usepackage[bbgreekl]{mathbbol}

\def\dfrac{\displaystyle\frac}

\renewcommand{\Im}{\mathop{\rm Im}}

\newcommand{\eps}{\varepsilon}
\renewcommand{\phi}{\varphi}

\newcommand{\FS}{\left|FS\right\rangle}

\begin{document}

\title{Magnetic-field control of Fermi polaron fine structure and polarization in strained monolayer semiconductors}

\author{Zakhar A. Iakovlev}
\affiliation{Ioffe Institute, 194021, St. Petersburg, Russia}

\begin{abstract}
    The theory of attractive Fermi polaron energy spectrum fine structure and polarization in doped two-dimensional semiconductors in external magnetic field is developed. Fermi polaron $g$ factor renormalization due to correlations with the valley-polarized Fermi sea of resident charge carriers is calculated. We study the competition between Zeeman and strain-induced splittings in monolayers under uniaxial strain. The control of strain, magnetic field and electron density allows continuous tuning of energy splitting and eigenstate polarization. Linearly polarized strain-induced split doublet exhibits quadratic Zeeman shift changing to linear Zeeman splitting with elliptical polarization with the increase of magnetic field. We identify a critical magnetic field above which resident charge carriers become fully valley polarized and only one circularly polarized Fermi polaron state remains. Within the Green's function approach we calculate energy levels and Stokes parameters of attractive Fermi polaron states and introduce a simplified effective two-level model allowing us to study analytically the interplay of Zeeman and pseudo-Zeeman splitting. We calculate absorption and reflection spectra, including circular and linear dichroism in the trion spectral range. These results show that real and strain-induced pseudomagnetic fields provide complementary tools for controlling the optical response of many-body excitonic quasiparticles in two-dimensional semiconductors.
\end{abstract}

\maketitle


\section{Introduction}
\label{sec:intro}

Electronic excitations and Coulomb complexes play a critical role in optical properties of semiconductors and semiconductor nanostructures~\cite{hawrylakOpticalPropertiesTwodimensional1991,ivchenko225,klingshirnElectronsPeriodicCrystal2007}. In recent years Coulomb complexes under study include both two-particle excitons and many-body complexes such as trions, biexcitons and even complexes with up to eight particles which are especially pronounced in low-dimensional systems, where Coulomb interaction is strongly enhanced~\cite{makTightlyBoundTrions2013,safwanExcitonsTrionsSemiconductor2015,baruaManyBodyExcitonInteractions2026,vantuanSixBodyEightBodyExciton2022,klenovskyCoulombCorrelatedMultiparticle2026}. The key motivation for investigation of many-body Coulomb complexes is the ability to control their energy spectrum fine structure and polarization, optical and transport properties. 

A powerful tool to study electronic systems is the external magnetic field. The Zeeman effect for charged excitons (trions), the three-particle bound states composed of an exciton with an electron or hole, was investigated in quantum wells, quantum dots and perovskites~\cite{astakhovBindingEnergyCharged2002,bayerFineStructureNeutral2002,kossackiPhotoluminescencePdopedQuantum2004,gladysiewiczEffectFreeCarriers2006,oberliCoulombCorrelationsCharged2009,rudno-rudzinskiMagnetoOpticalCharacterizationTrions2021b,huynhTransientQuantumBeatings2022}. Several theoretical works revisiting trions in magnetic field have appeared during the last several years~\cite{kleinTrionsAre2022,aleksandrovTwodimensionalTrionMagnetic2024,kudlisTheoryMagnetotrionpolaritonsTransition2024,jainExcitedStateTrionsQuantum2025}. The remarkable materials for investigation of many-body Coulomb complexes are transition metal dichalcogenide (TMDC) monolayers. TMDC monolayers host trions with large binding energy of $20$--$30$~meV due to the strong electron confinement in the monolayer plane and reduced dielectric screening, allow controlling the doping level via gating and provide a playground for studying the transport of Coulomb complexes~\cite{makTightlyBoundTrions2013,berkelbachTheoryNeutralCharged2013,ivchenko314,wagnerDiffusionExcitonsTwoDimensional2023a}.

In TMDC monolayers spin Zeeman and valley Zeeman effects were observed for neutral and charged excitons, dark excitons and trions, moir\'e exciton- and trion-polaritons~\cite{srivastavaValleyZeemanEffect2015,macneillBreakingValleyDegeneracy2015,aivazianMagneticControlValley2015,ivchenko297,deilmannInitioStudiesExciton2020,liValleySplittingPolarization2014a,liuGateTunableDark2019,zinkiewiczNeutralChargedDark2020,scherzerCorrelatedMagnetismMoire2024,lyonsGiantEffectiveZeeman2022,feuerIdentificationExcitonComplexes2023}. In early works about Zeeman effect for Coulomb complexes in TMDC monolayers the Zeeman splittings of excitons and trions were claimed to be the same. However, it was later shown that the correlations of trions with the Fermi sea of resident charge carriers should be taken into account~\cite{efimkinManybodyTheoryTrion2017,efimkinExcitonpolaronsDopedSemiconductors2018,massignanPolaronsAtomicGases2026}. It results in the different $g$ factors of trion states observed in experiments~\cite{lyonsValleyZeemanEffect2019,plechingerExcitonicValleyEffects2016a}. One of our goals is to calculate the renormalization of charged excitons $g$ factor within Fermi polaron (Suris tetron) model~\cite{glazovOpticalPropertiesCharged2020}, that takes into account correlations between trion and a Fermi sea of resident charge carriers. The main difference from the trion approach is that a tetron effectively consists of four particles: an electron in conduction band and a hole in valence band, constituting the exciton, and a Fermi sea electron-hole pair. The state  vacated by the excited electron in the conduction band appears to be a Fermi-hole. 

Another efficient control knob for manipulation of Coulomb complexes energy, polarization and fine structure is elastic strain~\cite{pengStrainEngineering2D2020}. In contrast to biaxial strain, that only shifts quasiparticle energies, uniaxial strain decreases the symmetry of the crystals, modifies electronic states and optical spectra, and is actively studied in various two-dimensional materials~\cite{schmidtReversibleUniaxialStrain2016,nicholsonUniaxialStraininducedPhase2021,niehuesUniaxialStrainTuning2022,aliRoomTemperaturePolarizationresolved2024,yutomoUniaxialStrainEffects2025,grimbergAngularEmissionProperties2026,evangelistaEffectsUniaxialStrain2026,chenAnomalousValleyHall2026}. Uniaxial strain splits the excitonic doublets into states linearly polarized along and perpendicular to the principal strain axes, while bare trions remain Kramers degenerate~\cite{finkelsteinNegativelyPositivelyCharged1996,astakhovBindingEnergyCharged2002,bayerFineStructureNeutral2002,glazovExcitonFineStructure2014,glazovExcitonFineStructure2022,jasinskiStrainInducedLifting2022}. In the previous works we showed that when taking into account correlations with the Fermi sea attractive (trion-like) Fermi polaron states, as they have integer spin, also split into linearly polarized components~\cite{iakovlevFermiPolaronFine2023,yagodkinFermiPolaronsStraininduced2025}.

Magnetic field and uniaxial strain provide two independent tools for controlling Fermi polaron states, motivating the study of their simultaneous action~\cite{blundoStrainInducedExcitonHybridization2022,andreevControllableFusionElectromagnetic2024}. The combination of magnetic field and uniaxial strain allows one to continuously control Fermi polaron states energies, their polarizations and energy spectrum fine structure splittings. The coaction of circularly-polarized Zeeman splitting and linearly polarized strain-induced splitting opens up an opportunity to construct and manage the Fermi polaron state with tunable elliptical polarization. This motivates the main goal of our paper: to develop a theory of the interplay between magnetic field-induced (Zeeman) and strain-induced (pseudo-Zeeman) splittings of Fermi polarons in doped TMDC monolayers.

The paper is organized as follows: after the introduction 
we present a model of Fermi polaron in an external magnetic field in doped W-based and Mo-based TMDC monolayers and optical properties of TMDC monolayer in Sec.~\ref{sec:model}. In Sec.~\ref{sec:zeeman} we develop a theory of Fermi polaron Zeeman effect. 
The interplay of Zeeman and strain-induced splittings is analyzed in Sec.~\ref{sec:strain}. 
The optical properties of TMDC monolayers including circular and linear dichroism of absorption and reflection in the attractive Fermi polaron spectral range are analyzed. We discuss the results in Sec.~\ref{sec:discussion} and present a summary of key results in Sec.~\ref{sec:conclusion}.

\section{Model}
\label{sec:model}

\begin{figure}[ht]
    \centering
    \includegraphics[width=\linewidth]{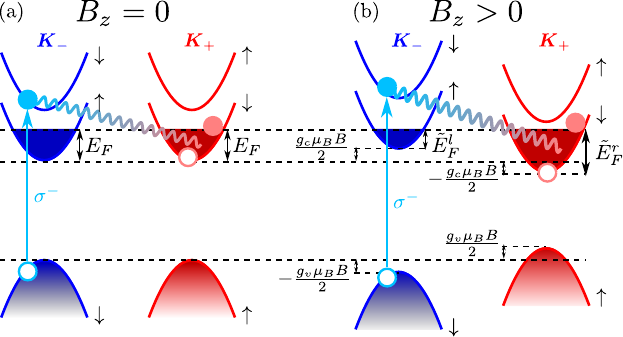}
    \caption{Intervalley Fermi polaron in W-based TMDC monolayer (a) without magnetic field and (b) in external magnetic field. Zeeman shifts of the bands, Eqs.~\eqref{eq:epsW}, are shown in panel (b). Exciton in $\bm K_-$-valley is excited by left circularly polarized light. Black arrows demonstrate electron spin in the subbands. The valley polarization is demonstrated via the change of Fermi energy, Eqs.~\eqref{eq:EF}, \eqref{eq:EFa}.}
    \label{fig:model}
\end{figure}

We start with a band structure model in doped TMDC monolayer. There are two valleys, $\bm K_+$ and $\bm K_-$, connected by time-reversal symmetry. The spin-orbit splitting exceeds $100$~meV in the valence band, allowing us to retain only topmost valence subband, whereas it is smaller in the conduction band $\sim 10$~meV. For the sake of certainty we consider negatively doped TMDC monolayer with the Fermi sea of electrons with relatively small electron density, for which the electrons partly fill only the lower conduction subband, see Fig.~\ref{fig:model}a. The energy scales in this paper are
\begin{equation}
    \label{eq:energy_scales}
    E_g \gg E_X \gg E_{T1,2}, |E| \gg |E_{T1} - E_{T2}|, \mu_BB, \hbar\Omega_X, E_F
\end{equation}
where $E_g$ is the band gap energy, $E_X$ is exciton binding energy, $E_{T1,2}$ are the binding energies of intra- and intervalley trions in W-based TMDC, $E$ is the difference between Coulomb complexes energy and exciton binding energy, $\mu_B$ is Bohr magneton, $B$ is magnetic field, $\hbar\Omega_X$ is characteristic strain energy described in the end of this section and $E_F$ is Fermi energy described below in Sec.~\ref{sec:FP}.

We introduce the Hamiltonian of interacting excitons and electrons in TMDC monolayer. Without external magnetic field the Hamiltonian of bare non-interacting electrons and optically active excitons~\cite{iakovlevFermiPolaronFine2023} is
\begin{equation}
    \label{eq:H0}
    \hat{\mathcal{H}}_0 = \sum_{\bm K}\eps^X_{\bm K}\left(\hat R^\dagger_{\bm K}\hat R_{\bm K} + \hat L^\dagger_{\bm K}\hat L_{\bm K}\right) + \sum_{\bm k}\eps_{\bm k}\left(\hat r^\dagger_{\bm k}\hat r_{\bm k} + \hat l^\dagger_{\bm k}\hat l_{\bm k}\right),
\end{equation}
where $\eps^X_{\bm K} = \hbar^2K^2/(2M_X)$ is exciton kinetic energy ($\bm K$ is exciton wavevector, $M_X = M_e + M_h$ is an exciton mass, $M_e$ is an electron mass and $M_h$ is a hole mass), $\eps_{\bm k} = \hbar^2k^2/(2M_e)$ is the kinetic energy of an electron in the bottom conduction subband ($\bm k$ is electron wavevector with respect to the valley center). Here, $\hat R^\dagger_{\bm K}$, $\hat r^\dagger_{\bm k}$, $\hat R_{\bm K}$ and $\hat r_{\bm k}$ are creation and annihilation operators of excitons and electrons in $\bm K_+$-valley, $\hat L^\dagger_{\bm K}$, $\hat l^\dagger_{\bm k}$, $\hat L_{\bm K}$ and $\hat l_{\bm k}$ are in $\bm K_-$-valley, respectively. Capital letters correspond to excitons with chiral selection rules, lowercase letters to electrons. We recall that in W-based TMDC monolayers optically bright excitons are excited in upper conduction subband, while in Mo-based TMDC monolayers electron forming bright exciton is excited in bottom conduction subband~\cite{deryPolarizationAnalysisExcitons2015,schneiderTwodimensionalSemiconductorsRegime2018,luMagneticFieldMixing2019}. The exciton-electron interaction Hamiltonian is~\cite{surisCorrelationTrionHole2003,glazovOpticalPropertiesCharged2020,iakovlevFermiPolaronFine2023,iakovlevLongitudinaltransverseSplittingFine2024}
\begin{multline}
    \hat{\mathcal{H}}_{int} = V_1\sum_{\bm k, \bm p, \bm p'}\left(\hat R^\dagger_{\bm k + \bm p - \bm p'}\hat r^\dagger_{\bm p'}\hat{R}_{\bm k}\hat{r}_{\bm p} + \hat L^\dagger_{\bm k + \bm p - \bm p'}\hat l^\dagger_{\bm p'}\hat{L}_{\bm k}\hat{l}_{\bm p}\right) \\ + V_2\sum_{\bm k, \bm p, \bm p'}\left(\hat R^\dagger_{\bm k + \bm p - \bm p'}\hat l^\dagger_{\bm p'}\hat{R}_{\bm k}\hat{l}_{\bm p} + \hat L^\dagger_{\bm k + \bm p - \bm p'}\hat r^\dagger_{\bm p'}\hat{L}_{\bm k}\hat{r}_{\bm p}\right),
\end{multline}
where the momentum conservation is preserved. $V_{1,2}$ are the intra- and intervalley interaction constants and can be expressed via trion energies, see below. For Mo-based systems due to the electron exchange interaction $V_1 > 0$, intravalley interaction is repulsive and only intervalley trions are observed. In contrast, in W-based TMDCs $|V_1 - V_2| \ll |V_{1,2}|$ and both intra- and intervalley trions are relevant.

\subsection{Coulomb complexes in external magnetic field}

In the presence of the static magnetic field perpendicular to the monolayer plane ($\bm B \equiv B\bm e_z$, where $z$ is a plane normal and $\bm e_z$ is the unit vector) the conduction and valence subbands have Zeeman shifts, Fig.~\ref{fig:model}b, resulting in the Zeeman splittings of exciton and electron energies (calculated relative to the zero-field case). We neglect orbital effects of the magnetic field due to the small electron scattering time $\tau_e$ ($|eB/M_ec|\tau_e \ll 1$), and retain only the Zeeman shifts of the relevant bands. We include the energy shifts in electron energies to take into account the effects of magnetic field. 

Due to the different order of conduction subbands the Zeeman shift of the lower sublevels is different for the W-based and Mo-based TMDC~\cite{luMagneticFieldMixing2019,koperskiOrbitalSpinValley2018}
\begin{subequations}
    \begin{align}
        \eps^{r,l}_{\bm k} = \eps_{\bm k} \mp \frac{g_c\mu_BB}{2}, \qquad & \text{W-based}, \\
        \label{eq:electronEMo}
        \eps^{r,l}_{\bm k} = \eps_{\bm k} \pm \frac{g_c\mu_BB}{2}, \qquad & \text{Mo-based},
    \end{align}
    where top sign corresponds to the $r$ index ($\bm K_+$-valley), for both types of systems $g_c$ is the $g$ factor of bottom conduction subband.
    
    The Zeeman shift of the exciton energy has a similar form both for Mo-based and W-based TMDCs~\cite{macneillBreakingValleyDegeneracy2015,koperskiOrbitalSpinValley2018}
    \label{eq:epsW}
    \begin{equation}
        \eps^R_{\bm K} = \eps^X_{\bm K} + \frac{g_X\mu_BB}{2}, \qquad \eps^L_{\bm K} = \eps^X_{\bm K} - \frac{g_X\mu_BB}{2},
    \end{equation}
    where $\eps^R_{\bm K}$ is the $\sigma^+$ polarized exciton energy and $\eps^L_{\bm K}$ refers to $\sigma^-$ polarized exciton. We remind that at our assumptions $\eps_0 = 0$. Here exciton $g$ factor
    \begin{align}
        g_X = g_{c'} - g_v, \qquad & \text{W-based}, \\
        g_X = g_c - g_v, \qquad & \text{Mo-based},
    \end{align}
    is the difference of the $g$ factors of the exciton components: the electron in the optically active conduction subband ($g_{c'}$ for upper subband in W-based TMDC and $g_c$ for lower conduction subband in Mo-based TMDC) and the electron in valence band ($g_v$ in electron representation), see Fig.~\ref{fig:model}.
\end{subequations}

Hence, the resulting Hamiltonian at $B \neq 0$ reads
\begin{multline}
    \hat{\mathcal{H}}_{B} = \sum_{\bm K}\left(\eps^R_{\bm K}\hat{R}^\dagger_{\bm K}\hat{R}_{\bm K} + \eps^L_{\bm K}\hat{L}^\dagger_{\bm K}\hat{L}_{\bm K}\right) \\ + \sum_{\bm k}\left(\eps^r_{\bm k}\hat{r}^\dagger_{\bm k}\hat{r}_{\bm k} + \eps^l_{\bm k}\hat{l}^\dagger_{\bm k}\hat{l}_{\bm k}\right).
\end{multline}
Combining with the interaction Hamiltonian and assuming that the field does not affect the parameters $V_1$ and $V_2$ we get the Hamiltonian of excitons interacting with the Fermi sea of doped TMDC monolayer in a form
\begin{equation}
    \label{eq:Hall}
    \hat{\mathcal{H}} = \hat{\mathcal{H}}_B + \hat{\mathcal{H}}_{int}.
\end{equation}
Here we focus on the Fermi polaron in the external magnetic field in isotropic TMDC monolayers, and provide a model for Fermi polaron Zeeman effect in uniaxially strained TMDC monolayers in the end of this section.

We find the eigenfunctions of this Hamiltonian: the bound Coulomb complexes. We mainly focus on the W-based systems where two trions: intra- and intervalley are optically active, and then extend results to the Mo-based systems with single attractive Fermi polaron branch. We solve the Schr\"odinger equation for the Hamiltonian~\eqref{eq:Hall}
\begin{equation}
    \label{eq:Schrodinger}
    \hat{\mathcal{H}}|\Psi_{\bm k}\rangle = E_{\bm k}|\Psi_{\bm k}\rangle,
\end{equation}
where we consider the quasiparticle with wavevector $\bm k$, wave function $|\Psi_{\bm k}\rangle$ and binding energy~$E_{\bm k}$. Here, we consider energy zero as the energy of bare exciton without interaction and external field. The energy of the particles in the unperturbed Fermi sea is also included in ground state energy.

In the absence of the uniaxial strain in Hamiltonian~\eqref{eq:Hall} the interaction is considered only between exciton and electrons, and excitons in $\bm K_+$- and in $\bm K_-$-valleys do not interact with each other. Therefore, we consider separately the complexes with excitons in $\bm K_+$-valley, that are excited in right circular polarization, and with excitons in $\bm K_-$-valley (excited by $\sigma^-$-polarized light). The wave function of the former is $|\Psi^R_{\bm k}\rangle$ and that of the latter is $|\Psi^L_{\bm k}\rangle$.

We study Coulomb complexes in the zero-temperature limit, in which the system ground state is characterized by the electrons in lower conduction subbands in both valleys below the same chemical potential. We determine this state as the state of unperturbed Fermi sea, $|FS\rangle$. To describe the bound quasiparticles in the low temperature and low exciton density limit, it is sufficient to consider the excitation of only one exciton together with at most one electron-hole pair in the Fermi sea. This quasiparticle is known as Fermi polaron or Suris tetron~\cite{surisExcitonsTrionsModified2001,surisCorrelationTrionHole2003,koudinovSurisTetronsPossible2014}.

The $\sigma^+$ polarized Fermi polaron wave function under the assumptions described above is
\begin{multline}
    \left|\Psi^R_{\bm k}\right\rangle = \phi_{\bm k}^R\hat{R}^\dagger_{\bm k}\FS + \sum_{\bm p, \bm q}F^{Rr}_{\bm k}(\bm p, \bm q)\hat{R}^\dagger_{\bm k - \bm p + \bm q}\hat{r}^\dagger_{\bm p}\hat{r}_{\bm q}\FS \\ + \sum_{\bm p, \bm q}F^{Rl}_{\bm k}(\bm p, \bm q)\hat{R}^\dagger_{\bm k - \bm p + \bm q}\hat{l}^\dagger_{\bm p}\hat{l}_{\bm q}\FS,
\end{multline}
and the similar form for $\sigma^-$ polarized state is obtained under the replacements $R \to L$ and $r \leftrightarrow l$. Further we use the notation $\sigma, \sigma' \in \{R, L\}$ for the exciton polarization and $\tau, \tau' \in \{r, l\}$ for the electron valley. In each particular equation the same symbols correspond to the same polarization/valley, while the symbols with primed indices describe the states in different polarizations/valleys ($\sigma \neq \sigma'$, $\tau \neq \tau'$). Note that $\sigma$ and $\tau$ may refer both to the same and to different valleys. Here, $\varphi^\sigma_{\bm k}$ is the bare excitonic amplitude of wave function ($|\varphi^\sigma_{\bm k}| \ll 1$ for attractive Fermi polarons and $|\varphi^\sigma_{\bm k}| \sim 1$ for repulsive Fermi polarons). $F^{\sigma\tau}_{\bm k}(\bm p, \bm q)$ and $F^{\sigma\tau'}_{\bm k}(\bm p, \bm q)$ are the amplitudes of states with the excitation of electron-hole pair in the Fermi sea in $\tau$ and $\tau'$ valley, respectively, $\bm p$ is the wavevector of excited electron, while $\bm q$ is the Fermi-hole wavevector.

In the absence of strain the Schr\"odinger equation~\eqref{eq:Schrodinger} is separated into two independent eigensystem problems for excitations in $\bm K_{\pm}$-valleys
\begin{subequations}
    \label{eq:HWall}
    \begin{equation}
        \hat{\mathcal{H}}|\Psi^\sigma_{\bm k}\rangle = E^\sigma_{\bm k}|\Psi^\sigma_{\bm k}\rangle.
    \end{equation}
    It gives us linear equations with the excitonic part
    \begin{multline}
        \eps^\sigma_{\bm k}\phi^\sigma_{\bm k} + V_{\sigma\tau}\sum_{\bm p, \bm q}F^{\sigma\tau}_{\bm k}(\bm p, \bm q) + V_{\sigma\tau'}\sum_{\bm p, \bm q}F^{\sigma\tau'}_{\bm k}(\bm p, \bm q) \\ + \sum_{\bm q}(V_1 + V_2)\varphi^\sigma_{\bm k} = E^\sigma_{\bm k}\varphi_{\bm k}^\sigma,
    \end{multline}
    and four-particle parts
    \begin{multline}
        \left(\eps^\sigma_{\bm k - \bm p + \bm q} + \eps^\tau_{\bm p} - \eps^\tau_{\bm q}\right)F^{\sigma\tau}_{\bm k}(\bm p, \bm q) + V_{\sigma\tau}\sum_{\bm p'}F^{\sigma\tau}_{\bm k}(\bm p', \bm q) \\ - V_{\sigma\tau}\sum_{\bm q'}F^{\sigma\tau}_{\bm k}(\bm p, \bm q') + V_{\sigma\tau}\phi^\sigma_{\bm k} = E^\sigma_{\bm k}F^{\sigma\tau}_{\bm k}(\bm p, \bm q),
    \end{multline}
    where from this point forward in summation we consider the summations over $\bm q$, $\bm q'$ are taken below chemical potential $\mu$ ($\eps^{r,l}_{\bm q} \leq \mu$) and summations over $\bm p$, $\bm p'$ are taken above Fermi level ($\eps^{r,l}_{\bm p} > \mu$). The valley in the summation is determined by the function, in which the summation argument appears, the valley of electron-Fermi-hole pair. The interaction parameters are $V_{Rr} = V_{Ll} \equiv V_1$ and $V_{Rl} = V_{Lr} \equiv V_2$. We remind, that we neglect the overall energy of unperturbed crystal due to nonzero electron density and magnetic field ($\left\langle FS\right|\mathcal{H}\FS$). In this paper we are interested in the trion-like attractive Fermi polarons, thus, we are looking for the states with the energy $E^\sigma_{\bm k} \approx -E_T$.
\end{subequations}

The system of Eqs.~\eqref{eq:HWall} gives us the self-consistent equation for the Fermi polaron energy
\begin{equation}
    \label{eq:EFP}
    E^\sigma_{\bm k} = \eps^\sigma_{\bm k} + \Sigma^\sigma_{\text{W},\bm k}(E^\sigma_{\bm k}),
\end{equation}
where we introduce the exciton self-energy
\begin{equation}
    \label{eq:SigmaW}
    \Sigma^\sigma_{\text{W},\bm k}\left(E^\sigma_{\bm k}\right) = \Sigma^{\sigma\tau}_{\bm k}\left(E^\sigma_{\bm k}\right) + \Sigma^{\sigma\tau'}_{\bm k}\left(E^\sigma_{\bm k}\right), 
\end{equation}
consisting of two summands, that separately describe the interaction with Fermi sea of electrons in each valley
\begin{equation}
    \label{eq:SigmaAa}
    \Sigma^{\sigma\tau}_{\bm k}\left(E^\sigma_{\bm k}\right) = \sum_{\bm q}\frac{V_{\sigma\tau}}{1 - V_{\sigma\tau}S^{\sigma\tau}_{\bm k}(\bm q)},
\end{equation}
where we also introduce the function
\begin{equation}
    \label{Eq:SAa}
    S^{\sigma\tau}_{\bm k}(\bm q) = \sum_{\bm p}\frac{1}{E^\sigma_{\bm k} - \left(\eps^\sigma_{\bm k - \bm p + \bm q} + \eps^\tau_{\bm p} - \eps^\tau_{\bm q}\right)}.
\end{equation}
In what follows, for simplicity we consider only the motionless Coulomb complexes (with $\bm k = 0$). We calculate energies of attractive Fermi polaron states and their optical properties in external magnetic field in Sec.~\ref{sec:zeeman}.

\subsection{Interplay of magnetic field and strain}

In strained monolayers the symmetry is reduced, circularly polarized states mix with each other in such a way that the eigenstates become linearly polarized along and across the main axis of strain and additional mixing may appear. We focus on the strain effects on excitons. Indeed, electrons and holes have half-integer spin and are Kramers degenerate. Strain may slightly affect the energy of the states, but does not lead to the fine structure. In contrast, excitons have integer spin and show strain-induced fine structure splitting~\cite{glazovExcitonFineStructure2022,iakovlevFermiPolaronFine2023}. In fact, excitons in $\bm K_+$- and $\bm K_-$-valleys form a two-level system and mix with each other due to the strain-induced Hamiltonian
\begin{equation}
    \label{eq:Hs}
    \hat{\mathcal{H}}_s = \sum_{\bm k}\frac{\hbar\Omega_X}{2}\left(\hat R^\dagger_{\bm k}\hat L_{\bm k} + \hat L^\dagger_{\bm k}\hat R_{\bm k}\right),
\end{equation}
where exciton energy splitting is linear in strain $\hbar\Omega_X \propto (u_{xx} - u_{yy})$ and $u_{\alpha\beta}$ are the strain tensor components in the strain main axes ($\alpha, \beta$ are Cartesian subscripts). We choose the direction of the main axes of strain in such a way that Eq.~\eqref{eq:Hs} takes this simple form. The typical value of strain-induced splitting~$|\hbar\Omega_X| \lesssim 1$~meV, that is less or of the same order as $E_F$, $\mu_BB$ and $|E_{T1} - E_{T2}|$, Eq.~\eqref{eq:energy_scales}.


The Hamiltonian of the Coulomb complexes in strained TMDC monolayers in perpendicular magnetic field is
\begin{equation}
    \hat{\mathcal{H}} = \hat{\mathcal{H}}_B + \hat{\mathcal{H}}_{int} + \hat{\mathcal{H}}_s.
\end{equation}
It changes the Schr\"odinger equation~\eqref{eq:HWall} and mixes all four attractive Fermi polaron states (in W-based TMDC) due to the exciton valley mixing~\eqref{eq:Hs}
\begin{subequations}
    \begin{multline}
        \eps^\sigma_{\bm k}\phi^\sigma_{\bm k} + V_{\sigma\tau}\sum_{\bm p, \bm q}F^{\sigma\tau}_{\bm k}(\bm p, \bm q) + V_{\sigma\tau'}\sum_{\bm p, \bm q}F^{\sigma\tau'}_{\bm k}(\bm p, \bm q) \\ + \sum_{\bm q}(V_1 + V_2)\varphi^\sigma_{\bm k} + \frac{\hbar \Omega_X}{2}\varphi^{\sigma'}_{\bm k} = E_{\bm k}\varphi_{\bm k}^\sigma,
    \end{multline}
    \begin{multline}
        \left(\eps^\sigma_{\bm k - \bm p + \bm q} + \eps^\tau_{\bm p} - \eps^\tau_{\bm q}\right)F^{\sigma\tau}_{\bm k}(\bm p, \bm q) \\ + V_{\sigma\tau}\sum_{\bm p'}F^{\sigma\tau}_{\bm k}(\bm p', \bm q) - V_{\sigma\tau}\sum_{\bm q'}F^{\sigma\tau}_{\bm k}(\bm p, \bm q') \\ + V_{\sigma\tau}\phi^\sigma_{\bm k} + \frac{\hbar\Omega_X}{2}F^{\sigma'\tau}_{\bm k}(\bm p, \bm q) = E_{\bm k}F^{\sigma\tau}_{\bm k}(\bm p, \bm q).
    \end{multline}
\end{subequations}

Due to the valley mixing, exciton self-energy~\eqref{eq:SigmaAa} is renormalized as
\begin{equation}
    \label{eq:SigmasAa}
    \Sigma^{\sigma\tau}_{\text s,\bm k}\left(E_{\bm k}\right) = \sum_{\bm q}\frac{\left[1 - \dfrac{V_{\sigma'\tau}V_{\sigma\tau}S_{\text s}^2}{1 - V_{\sigma'\tau}S^{\sigma'\tau}_{\bm k}(\bm q)}\right]V_{\sigma\tau}}{\left[1 - V_{\sigma\tau}S^{\sigma\tau}_{\bm k}(\bm q)\right] - \dfrac{V_{\sigma'\tau}V_{\sigma\tau}S_{\text s}^2}{1 - V_{\sigma'\tau}S^{\sigma'\tau}_{\bm k}(\bm q)}},
\end{equation}
where the bottom index `s', meaning strain, is introduced to distinguish this self-energy from the self-energy without strain, Eq.~\eqref{eq:SigmaAa}. We introduce the mixing parameter $\propto \hbar\Omega_X/E_T$ independent on $\bm q$ in its leading term
\begin{equation}
    S_{\text s} = \sum_{\bm p}\frac{\hbar\Omega_X/2}{\left[E_{\bm k}- \left(\eps^X_{\bm k - \bm p + \bm q} + \eps_{\bm p} - \eps_{\bm q}\right)\right]^2} \approx - \mathcal{D}\frac{\hbar\Omega_X}{2E_{\bm k}}.
\end{equation}
Note, that the second term in numerator in Eq.~\eqref{eq:SigmasAa} is beyond the validity of the theory and should be neglected due to the factor of $\hbar\Omega_X/\left[E_T\ln\left(E_X/E_T\right)\right] \ll 1$.

Now, Fermi polaron states are coupled and self-consistent equations for Fermi polaron energy become quadratic equations
\begin{multline}
    \label{eq:FPsEnergy}
    \left[E_{\bm k} - \eps^R_{\bm k} - \Sigma^R_{\text s, \bm k}(E_{\bm k})\right]\left[E_{\bm k} - \eps^L_{\bm k} - \Sigma^L_{\text s, \bm k}(E_{\bm k})\right] \\ = \Xi^R_{\bm k}(E_{\bm k})\Xi^L_{\bm k}(E_{\bm k}),
\end{multline}
where as in Eq.~\eqref{eq:SigmaW}
\begin{equation}
    \Sigma^\sigma_{\text s,\bm k}\left(E_{\bm k}\right) = \Sigma^{\sigma\tau}_{\text s,\bm k}\left(E_{\bm k}\right) + \Sigma^{\sigma\tau'}_{\text s,\bm k}\left(E_{\bm k}\right)
\end{equation}
and mixing-induced terms $\propto \hbar\Omega_X$ are
\begin{equation}
    \Xi^\sigma_{\bm k}(E_{\bm k}) = \frac{\hbar\Omega_X}{2} + \Xi^{\sigma\tau}_{\bm k}(E_{\bm k}) + \Xi^{\sigma\tau'}_{\bm k}(E_{\bm k}),
\end{equation}
with each particular term
\begin{equation}
    \Xi^{\sigma\tau}_{\bm k}(E_{\bm k}) = \sum_{\bm q}\frac{\dfrac{V_{\sigma'\tau}V_{\sigma\tau}S_{\text s}}{1 - V_{\sigma'\tau}S^{\sigma'\tau}_{\bm k}(\bm q)}V_{\sigma\tau}S_{\bm k}^{\sigma\tau}(\bm q)}{\left[1 - V_{\sigma\tau}S^{\sigma\tau}_{\bm k}(\bm q)\right] - \dfrac{V_{\sigma'\tau}V_{\sigma\tau}S_{\text s}^2}{1 - V_{\sigma'\tau}S^{\sigma'\tau}_{\bm k}(\bm q)}}.
\end{equation}
Without strain energy levels decouple and Fermi polarons with excitons in $\bm K_{\pm}$-valleys are independent. 

In contrast, without external magnetic field ($B = 0$) $R$ and $L$ indices are equivalent, and eigenstates are polarized along and across the strain direction. In the limit of large $T_1$~-- $T_2$ splitting
\begin{equation}
    \label{eq:small}
    |E_{T1} - E_{T2}| \gg E_F, \hbar\Omega_X, |\mu_BB|,
\end{equation}
the splitting takes form
\begin{equation}
    \label{eq:DEFPs}
    \Delta E_{\text{FP}}^{(s)} = \frac{\alpha E_F\hbar\Omega_X}{\sqrt{(E_{T1} - E_{T2})^2/4 + (\alpha E_F)^2}},
\end{equation}
as was predicted in~\cite{iakovlevFermiPolaronFine2023}. The numerical coefficient $\alpha$ is defined below in Eq.~\eqref{eq:alpha}. Note that the additional term $\propto E_F^2$ exceeds the accuracy, but we keep it for the better agreement with numerical calculations.

The present model extends the results of the previous subsection and paper~\cite{glazovOpticalPropertiesCharged2020} on the strained samples and the results of papers~\cite{iakovlevFermiPolaronFine2023,yagodkinFermiPolaronsStraininduced2025} on the application of magnetic field. The Fermi polaron energies and oprical properties of strained TMDC monolayers are presented in Sec.~\ref{sec:strain}.

The proposed approach allows us to calculate the Fermi polaron energy, its Green's function and polarization of states. Further, to calculate the absolute value of absorption and reflection together with their linear and circular dichroism in external magnetic field and under strain, we need to express the optical properties of TMDC monolayers in terms of Fermi polaron Green's function.

\subsection{Optical properties}

We derive the optical properties of a TMDC monolayer in the energy range of attractive Fermi polaron under normal incidence of light. The monolayer polarization~$\bm P(z)$ is determined by the in-plane electric field component~$\bm E_\parallel(z)$ and the resonant $2\times 2$ tensor Green's function of the Coulomb complexes in TMDC~$\hat{\mathcal{G}}(\omega)$. For normal incidence of light
\begin{equation}
    \bm P(z) = -\frac{q\,\delta(z)}{2\pi q_0^2}\hbar\Gamma_0\hat{\mathcal{G}}(\omega)\bm E_\parallel(z),
\end{equation}
where $q_0 = \omega/c$ is the wavevector of light in vacuum, $q = \sqrt{\varkappa}q_0$ is the wavevector of light in the background medium ($\varkappa$ is its dielectric constant) and $\Gamma_0 = 1/(2\tau)$ is the exciton radiative decay rate ($\tau$ is exciton radiative lifetime)~\cite{ivchenko225, glazovExcitonFineStructure2014, prazdnichnykhControlExcitonValley2021}. Note that anisotropic tensor $\hat{\mathcal{G}}(\omega)$ results in the different polarizations of electric field $\bm E_\parallel(z)$ and polarization $\bm P(z)$. The Dirac $\delta$-function signifies that the polarization is localized in the monolayer.

Under normal incidence of light $\bm E_\parallel(z) = \bm E(z)$ and the wave equation
\begin{equation}
    \frac{d^2\bm E(z)}{dz^2}+q^2\bm E(z) = -q_0^2\,4\pi\bm P(z)
\end{equation}
together with the continuity of the magnetic field gives matrix expressions for the reflection coefficient
\begin{equation}
    \label{eq:r}
    \hat{r} = \left[\hat{1} + {\rm i}\hbar\Gamma_0\hat{\mathcal{G}}(\omega)\right]^{-1}\left[-{\rm i}\hbar\Gamma_0\hat{\mathcal{G}}(\omega)\right]
\end{equation}
and transmission coefficient
\begin{equation}
    \label{eq:t}
    \hat{t} = \hat{1} + \hat{r},
\end{equation}
where $\hat 1$ is $2\times 2$ unity tensor. In this paper we only consider optical properties of TMDC monolayers in vacuum. In encapsulated samples reflection and transmission coefficients should be renormalized in terms of the refractive index of surrounding medium~\cite{prazdnichnykhControlExcitonValley2021}.

For the scalar (isotropic) Green's function this expression reduces to the standard result
\begin{equation}
    r = \frac{-{\rm i}\hbar\Gamma_0}{\mathcal{G}^{-1}(\omega) + {\rm i}\hbar\Gamma_0}.
\end{equation}
This result is also valid for the excitation along the main axis of tensor $\hat{\mathcal{G}}(\omega)$ with the principal value of the tensor $\hat{\mathcal{G}}(\omega)$ along the corresponding axis.

The absorption of light by TMDC monolayer~$\mathcal{A}$ can be found as a complement to the total reflection~$\mathcal{R} = |\hat r \bm e|^2$ and transmission~$\mathcal{T} = |\hat t\bm e|^2$,where $\bm e$ is the polarization of incident light,
\begin{equation}
    \label{eq:As}
    \mathcal{A} = 1 - \mathcal{R} - \mathcal{T}.
\end{equation}

In the isotropic case for sufficiently small radiative decay rate $\hbar\Gamma_0\mathcal{G}(\omega) \ll 1$ we get the known result $\mathcal{A} \propto -\text{Im}\left[\mathcal G(\omega)\right]$.


\section{Results}

In this section we start from the study of the Zeeman effect without uniaxial strain and then focus on the interplay of the external magnetic field and uniaxial strain.

\subsection{Zeeman effect}
\label{sec:zeeman}

Before developing a theory of Fermi polaron Zeeman effect, we start with the bare trion.

\subsubsection{Trion}

The trion energy is found from the condition~\cite{surisCorrelationTrionHole2003,glazovOpticalPropertiesCharged2020}
\begin{equation}
    V_{\sigma\tau}S^{\sigma\tau}_{\bm k}(0) = 1.
\end{equation}

In the absence of magnetic field, the trion binding energy is determined by the exciton energy
\begin{equation}
    E_{T1,2} = E_X\exp\left(\frac{1}{\mathcal{D}V_{1,2}}\right),
\end{equation}
where $\mathcal{D} = (M_eM_X)/(2\pi\hbar^2M_T)$ is the exciton-electron reduced density of states ($M_T = M_e + M_X$ is the trion mass), and $E_X$ plays a role of cutoff energy in summation over $\bm p$ in Eq.~\eqref{Eq:SAa}. Note, that in all summations we omit the normalization area.

With an external magnetic field applied the additional Zeeman splitting appears. We consider magnetic field sufficiently weak and not changing the internal trion structure. The magnetic field simply shifts the energy of the particles forming a trion, and after the recombination electron in conduction band remains in the same subband, and its Zeeman shift ($\propto g_c$) does not affect trion binding energy. Therefore, the energy calculated with respect to exciton energy in the absence of magnetic field of the trion energy combined from the exciton in $\sigma$-valley and electron in $\tau$-valley is
\begin{equation}
    E^{\sigma\tau}_{T} = -E_{T,\sigma\tau} + \eps_0^\sigma,
\end{equation}
with the same convention, intravalley trion binding energy is $E_{T,Rr} = E_{T,Ll} \equiv E_{T1}$ and intervalley energy is $E_{T,Rl} = E_{T,Lr} \equiv E_{T2}$. 
Thus, the trion splitting
\begin{equation}
    \label{eq:Tbare}
    \Delta E_T = g_X\mu_BB
\end{equation}
is the same as for neutral exciton and for both trion states their $g$ factors are $g_T = g_X$.

\subsubsection{Fermi polaron}
\label{sec:FP}

We need to estimate the exciton self-energy depending on the total electron density $N_e$ accounting for both valleys. For brevity, we express all results in terms of Fermi energy
\begin{equation}
    \label{eq:EF}
    E_F = \frac{\pi\hbar^2}{M_e}N_e,
\end{equation}
that corresponds to the chemical potential $\mu$ without external magnetic field, Fig.~\ref{fig:model}a. Magnetic field shifts the energies of conduction bands, thus, there is a critical magnetic field
\begin{equation}
    B_{\text{crit}} = \frac{2E_F}{|g_c|\mu_B}.
\end{equation}
For sufficiently small magnetic field ($|B| < B_{\text{crit}}$) both valleys are filled with electrons below the initial chemical potential $\mu$ (as density of states in two-dimensional materials is constant), Fig.~\ref{fig:model}b. At the critical magnetic field resident electrons in conduction band appear to be fully polarized, and at large magnetic field ($|B| > B_{\text{crit}}$) all electrons are in the one valley. All quantities are determined by the electron density in a particular valley,
\begin{equation}
    \label{eq:EFa}
    \tilde{E}_F^\tau = \begin{cases}
        0, \quad & E_F < \eps_0^\tau, \\
        E_F - \eps_0^\tau, \quad & -E_F < \eps_0^\tau < E_F, \\
        2E_F, \quad & \eps_0^\tau < -E_F, \\
    \end{cases}
\end{equation}

The calculation of exciton self-energy $\Sigma^{\sigma\tau}_0\left(E^\sigma_0\right)$ in W-based TMDC monolayer is presented in Appendix~\ref{app:SigmaW}. Fermi polaron energy is the solution of self-consistent equation~\eqref{eq:EFP} at $\bm k = 0$. In the limit of small electron densities and magnetic fields, Eq.~\eqref{eq:small}, two attractive Fermi polaron states do not interfere with each other, and Fermi polaron energies are solutions of equations with one self-energy term (single trion pole approximation)
\begin{equation}
    \label{eq:EAa_single_pole}
    E_0^{\sigma\tau} = \eps_0^\sigma+\Sigma_0^{\sigma\tau}\left(E_0^{\sigma\tau}\right).
\end{equation}
It gives energy shift in a linear in $\tilde{E}_F^\tau$ approximation
\begin{equation}
    \label{eq:EFPW}
    E^{\sigma\tau}_{\text{FP}} = -E_{T,\sigma\tau} + \eps^\sigma_0 + \xi\tilde{E}_F^\tau,
\end{equation}
where the factor $\xi$ is determined only by the particle masses
\begin{equation}
    \xi = \frac{M_T}{M_X} - \frac{M_X/M_T}{1 - \exp\left[-\left(\frac{M_X}{M_T}\right)^2\right]}.
\end{equation}
For the equal electron and hole masses ($M_e = M_h$), $\xi = -0.36$. For arbitrary mass ratio $\xi$ is always negative, resulting in the increase of Fermi polaron binding energy $|E^{\sigma\tau}_{\text{FP}}|$ with the increase of electron density.

One can see that for attractive Fermi polarons not only the exciton splitting $\Delta E_T$ appears, but also the splitting, caused by the valley energy shift (the difference in $\tilde{E}_F^r$ and $\tilde{E}_F^l$). At small magnetic fields ($|B| < B_{\text{crit}}$) difference in density-induced energy shift appears due to the difference in the electron density. It affects both the Fermi polaron energy shift and oscillator strength $\propto \tilde{E}_F^\tau$~\cite{glazovOpticalPropertiesCharged2020}. Finally, the Fermi polaron energy splittings in magnetic field are
\begin{subequations}
\label{eq:dEWFP}
\begin{equation}
    \Delta E^{\text{W}}_{\text{FP}1,2} = g_{1,2}\mu_BB,
\end{equation}
where the Zeeman effect is different for intra- and intervalley Fermi polarons. For intravalley Fermi polaron
\begin{equation}
    g_1 = g_X + \xi g_c
\end{equation}
and for intervalley Fermi polaron
\begin{equation}
    g_2 = g_X - \xi g_c.
\end{equation}
\end{subequations}
This effect appears even at $E_F \to 0$, as soon as magnetic field is sufficiently small, which distinguish attractive Fermi polaron states from bare trion states. The different $g$ factors for different Coulomb-complex states in TMDC monolayers were previously observed~\cite{lyonsValleyZeemanEffect2019,plechingerExcitonicValleyEffects2016a}.

For sufficiently large magnetic fields ($|B| > B_{\text{crit}}$) only one valley contains electrons, thus, only Fermi polaron involving an electron from the occupied valley $\tau$ can be combined~\cite{backGiantParamagnetismInducedValley2017,smolenskiInteractionInducedShubnikovdeHaas2019}. The polaron-induced energy shift remains depending only on electron density, and magnetic field-induced energy shift is the same as for trion case with $g_{\text{FP}} = g_T = g_X$.

\begin{figure}
    \centering
    \includegraphics[width=\linewidth]{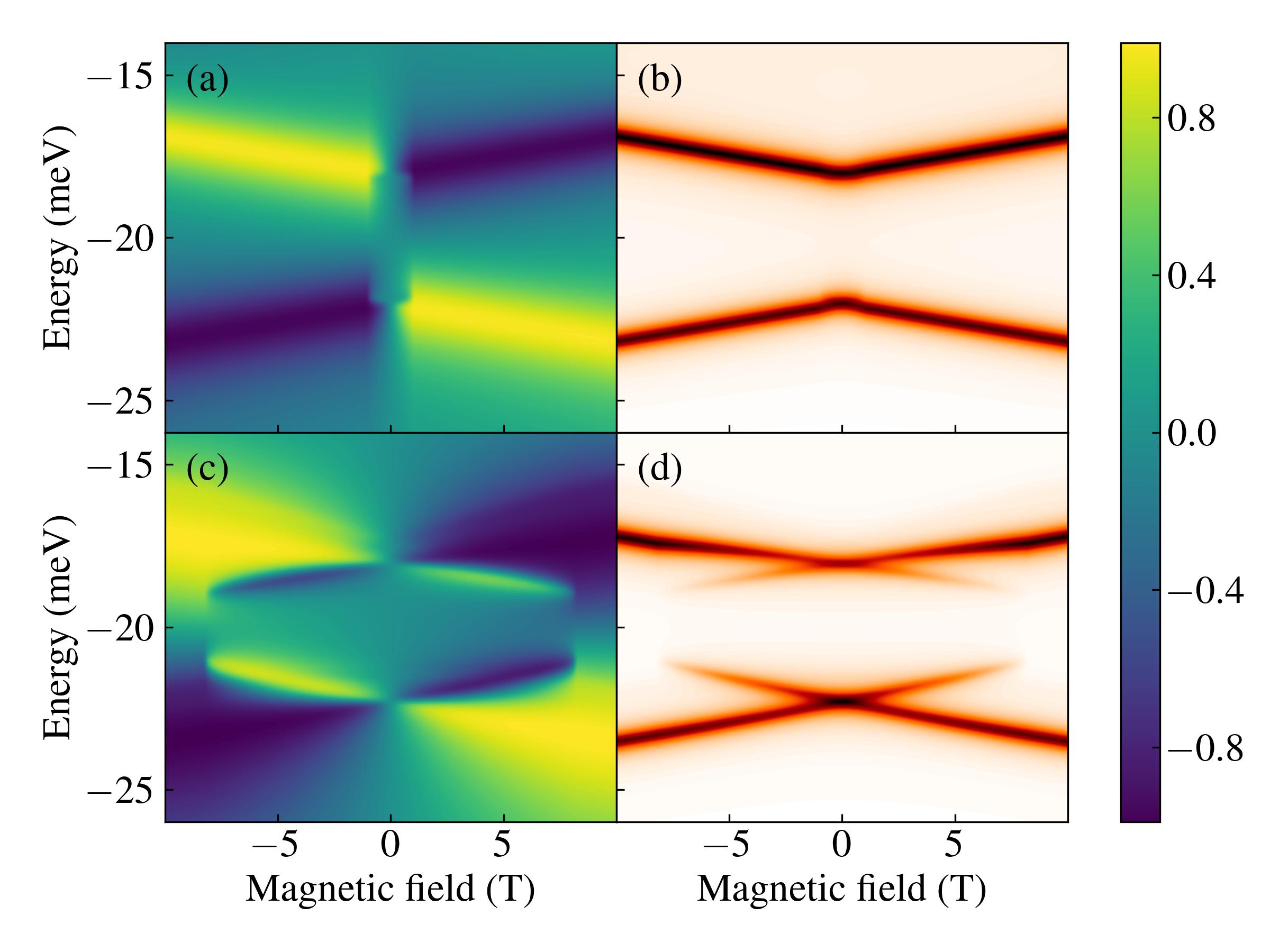}
    \caption{(a,c) Circular dichroism of absorption $P_c(E)$, Eq.~\eqref{eq:Pc}, and (b,d) absorption spectra of unpolarized light $\mathcal{A}$, Eq.~\eqref{eq:abs}, as a function of magnetic field for W-based TMDC. (a,b)~Small doping $N_e = 10^{10}$~cm$^{-2}$, $B_{\text{crit}} = 1$~T, (c,d)~moderate doping $N_e = 8\cdot 10^{10}$~cm$^{-2}$, $B_{\text{crit}}=8$~T. The parameters are $E_{T1} = 22$~meV, $E_{T2} = 18$~meV, $M_e = M_h = 0.4m_0$, $g_X = -4$, $g_c = 2$ and $\hbar\gamma = 0.2$~meV.}
    \label{fig:Pol(B)}
\end{figure}

The absorption spectra, Eq.~\eqref{eq:As} in $\sigma$-polarized circular polarization $\mathcal{A}^\sigma\propto -\Im\left[\mathcal{G}^\sigma(E + {\rm i}\hbar\gamma)\right]$ with phenomenological damping $\gamma$ are determined by Fermi polaron Green's function
\begin{equation}
    \mathcal{G}^\sigma(E) = \frac1{E - \eps_0^\sigma - \Sigma^\sigma_{\text{W},0}(E)}.
\end{equation}
Due to the difference in electron densities in $\bm K_{\pm}$-valleys, the oscillator strengths of attractive Fermi polaron states differ resulting in the circular dichroism of absorption
\begin{equation}
    \label{eq:Pc}
    P_c(E) = \frac{\mathcal{A}^R -\mathcal{A}^L}{\mathcal{A}},
\end{equation}
where the total absorption of unpolarized light is
\begin{equation}
    \label{eq:abs}
    \mathcal{A} = \mathcal{A}^R + \mathcal{A}^L.
\end{equation}
Circular dichroism of absorption~$P_c(E)$ and absorption spectra for different electron densities as a function of magnetic field are shown in Fig.~\ref{fig:Pol(B)}. The attractive Fermi polaron sublevels are split in external field. At small doping, Fig.~\ref{fig:Pol(B)}(a,b), electrons in conduction bands are completely polarized already at small magnetic field $|B| > B_{\text{crit}} = 1$~T, thus, only one sublevel of each attractive Fermi polaron doublet is observed with electron from filled valley. At moderate doping, Fig.~\ref{fig:Pol(B)}(c,d), the boundary of fully valley-polarized regime is clearly seen at $B_{\text{crit}} = 8$~T. At large magnetic field the same regime as at small doping is observed. At small magnetic field both circularly polarized attractive Fermi polaron states are observed with the redistribution of oscillator strengths between the sublevels with the increase of magnetic field due to the valley polarization. 

The numerical results and analytical derivations can be simplified using trion pole approximation for self-energies
\begin{equation}
    \Sigma^{\sigma\tau}_0\left(E^\sigma_0\right) = \frac{\alpha \tilde{E}^\tau_FE_{T,\sigma\tau}}{E^\sigma_0 - \eps^\sigma_0 + E_{T,\sigma\tau} - \beta\tilde{E}^\tau_F},
\end{equation}
where the coefficients
\begin{equation}
    \label{eq:alpha}
    \alpha = \frac{\left(M_X/M_T\right)^3}{4\sinh^2\left[\frac{1}{2}\left(\frac{M_X}{M_T}\right)^2\right]}, \quad \beta = \alpha + \xi,
\end{equation}
are determined in a such way, that in a linear in $\tilde E_F^\tau$ regime Fermi polaron energies and oscillator strengths are the same as derived from Eq.~\eqref{eq:EAa_single_pole}. For equal electron and hole masses, $M_e = M_h$, $\alpha = 1.48$ and $\beta = 1.12$.

Accounting for both intravalley and intervalley interactions in the trion-pole approximation gives the expression for the Fermi polaron energy, Eq.~\eqref{eq:EFPW}, beyond the approximation~\eqref{eq:small} in a general approximation~\eqref{eq:energy_scales} in a form
\begin{subequations}
    \label{eq:EW}
    \begin{multline}
        E^{R\tau}_0 = -E_T + \eps_0^R + \xi E_F \\ \pm \sqrt{\left[\frac{E_{T1} - E_{T2}}{2} + \frac{\xi}{2}\left(\tilde{E}^l_F - \tilde{E}^r_F\right)\right]^2 + \alpha^2\tilde{E}^l_F\tilde{E}^r_F},
    \end{multline}
    with ``$+$'' for intervalley Fermi polaron ($\tau=l$) and ``$-$'' for intravalley Fermi polaron ($\tau = r$). For the Fermi polaron excited with $\sigma^-$ polarized excitation the energy is obtained by making the replacement $R \leftrightarrow L$ and $r \leftrightarrow l$ again with ``+'' for intervalley Fermi polaron ($\tau = r$). 
\end{subequations}
Here, $E_T = (E_{T1} + E_{T2}) / 2$ is the average trion binding energy.

For Mo-based systems the order of spin sublevels in conduction band is opposite to the W-based TMDC, thus, the energy shifts are different in bottom conduction band, Eq.~\eqref{eq:electronEMo}. 
The intravalley exciton-electron interaction is repulsive ($V_1 > 0$), thus, only intervalley trions exist and are taken into account. Exciton self-energy in Mo-based TMDC is presented in Appendix~\ref{app:SigmaMo}. Attractive Fermi polaron energy in Mo-based TMDC is
\begin{equation}
    \label{eq:EMo}
    E^R_0 = -E_T+\eps_0^R + \xi \tilde{E}^l_F, \qquad E^L_0 = -E_T+\eps_0^L + \xi \tilde{E}^r_F,
\end{equation}
that coincides with the linear in $E_F$ and $B$ regime for W-based systems, Eq.~\eqref{eq:EFPW}. For small magnetic field the splitting is
\begin{equation}
    \label{eq:dEMoFP}
    \Delta E^{\text{Mo}}_{\text{FP}} = \Delta E_T + \xi g_c\mu_BB,
\end{equation}
that leads to $g$ factor $g_{\text{Mo}} = g_X + \xi g_c$, coinciding with the intravalley $g$ factor for W-based TMDC due to the same sign of relative bottom conduction subband energy shift with respect to the exciton Zeeman shift. 

\begin{figure}
    \centering
    \includegraphics[width=\linewidth]{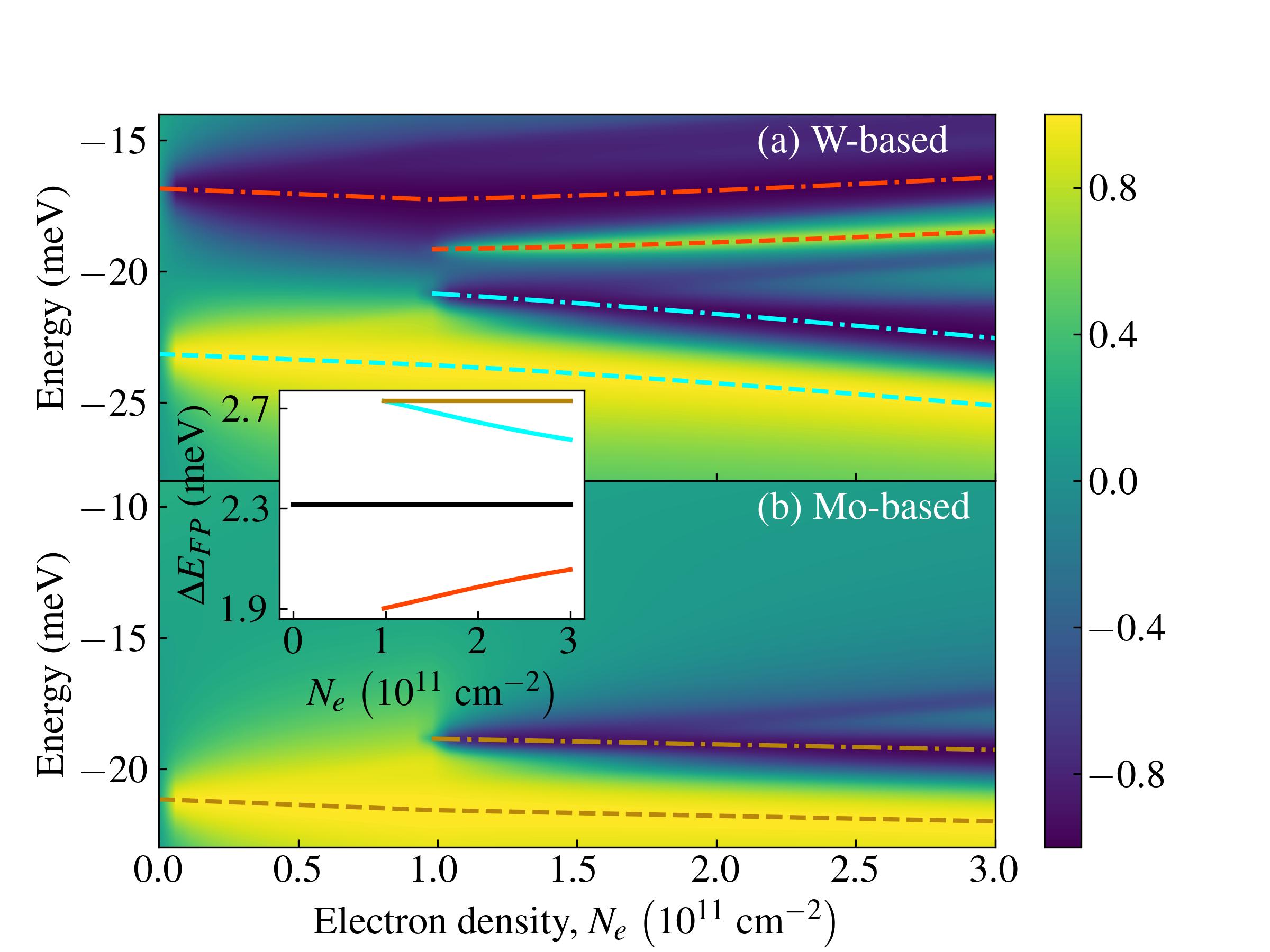}
    \caption{Circular dichroism of absorption $P_c(E)$, Eq.~\eqref{eq:Pc}, in the spectral range of attractive Fermi polarons at magnetic field $B = 10$~T for (a) W-based TMDC and (b) Mo-based TMDC. Red and cyan lines show inter- and intravalley Fermi polaron levels in W-based TMDC, Eq.~\eqref{eq:EW}, brown lines show intervalley Fermi polaron levels in Mo-based TMDC, Eq.~\eqref{eq:EMo}. Dashed lines show $R$-polarized sublevels, dashed-dotted lines show $L$-polarized sublevels. The inset demonstrates the attractive Fermi polaron energy splittings with the same colors as the main figure, black line shows exciton splitting, that is equal to the bare trion splitting. 
    The critical electron density, corresponding to $B_{\text{crit}} = 10$~T, is $N_e = 10^{11}$~cm$^{-2}$.
    The calculation parameters for W-based TMDC are the same as in Fig.~\ref{fig:Pol(B)}, for Mo-based TMDC $E_T = 20$~meV, and other parameters are the same.}
    \label{fig:Pol(N)}
\end{figure}

Circular dichroism of absorption $P_c(E)$ for different electron densities in an external magnetic field $B = 10$~T for W-based and Mo-based TMDC is shown in Fig.~\ref{fig:Pol(N)}. At $N_e \to 0$ conduction band is empty, and attractive Fermi polarons do not appear. At small electron densities, $E_F < |g_c\mu_BB|/2$ ($B > B_{\text{crit}}$), electrons are only in one valley, and only Fermi polarons with electron in $K$~valley in W-based TMDC and with electron in $K'$~valley in Mo-based TMDC are present. At critical electron density $N_e = 10^{11}$~cm$^{-2}$, corresponding to $B_{\text{crit}} = 10$~T, both valleys begin to be occupied with electrons, and all kinds of attractive Fermi polarons appear (in Fig.~\ref{fig:Pol(N)} only the observed states are shown). At this electron density energy levels kink. The observed attractive Fermi polaron Zeeman splitting is already different from the exciton Zeeman splitting (which is equal to the bare trion splitting), Eq.~\eqref{eq:Tbare}, Fig.~\ref{fig:Pol(N)} inset. In W-based TMDC the splitting further changes with the increase of electron density and tends to the splitting of free exciton, Eq.~\eqref{eq:EW}, due to the interaction with electrons in both valleys. In contrast, in Mo-based TMDC the interaction is only intervalley, and the Zeeman splitting does not change with the increase in electron density. 

\subsection{Strained TMDC monolayers}
\label{sec:strain}


The simultaneous application of external magnetic field and uniaxial strain allows full control of general elliptical Fermi polaron polarization. The attractive Fermi polaron Green's function has an essentially tensor nature both in circular and in Cartesian bases. In the circular basis, Fermi polaron Green's function takes form
\begin{equation}
    \label{eq:Gs}
    \hat{\mathcal G}_s(E) = \begin{pmatrix}
        \mathcal{G}^{RR}_s(E) & \mathcal{G}^{RL}_s(E) \\
        \mathcal{G}^{LR}_s(E) & \mathcal{G}^{LL}_s(E)
    \end{pmatrix},
\end{equation}
resulting in circular dichroism at $\hbar\Omega_X = 0$ and splitting into linearly polarized states at $B = 0$. The Green's function components in circular basis (for $\bm k = 0$) are
\begin{subequations}
    \begin{equation}
        \mathcal{G}^{\sigma\sigma}_s(E) = \frac{1}{E - \eps^\sigma_0 - \Sigma^\sigma_{\text s,0}(E) - \dfrac{\Xi^R_0(E)\Xi^L_0(E)}{E - \eps^{\sigma'}_0 - \Sigma^{\sigma'}_{\text s,0}(E)}},
    \end{equation}
    \begin{equation}
        \mathcal{G}^{\sigma\sigma'}_s(E) = \frac{\Xi^{\sigma'}_0(E) / \left[E - \eps^{\sigma'}_0 - \Sigma^{\sigma'}_{\text s,0}(E)\right]}{E - \eps^\sigma_0 - \Sigma^\sigma_{\text s,0}(E) - \dfrac{\Xi^R_0(E)\Xi^L_0(E)}{E - \eps^{\sigma'}_0 - \Sigma^{\sigma'}_{\text s,0}(E)}}.
    \end{equation}
\end{subequations}


The exciton self-energy $\Sigma^{\sigma\tau}_{\text s,0}(E)$ and mixing-induced term $\Xi^\sigma_0(E)$ in the assumptions of energy scales from Eq.~\eqref{eq:energy_scales} are derived in Appendix~\ref{app:SigmaSW}. In this limit the symmetry relations $\Xi^L_0(E) = \Xi^R_0(E) = \Xi(E)$ lead to the symmetry of Green's function, Eq.~\eqref{eq:Gs}, $\mathcal G^{RL}_s(E) = \mathcal G^{LR}_s(E)$.

To get a physical understanding we imagine the interplay of independent impacts of magnetic field and anisotropic strain and formulate a simplified analytical model considering intra- and intervalley Fermi polaron doublets as a two-level system. Generally, it is not the case, as valley polarization affects the oscillator strengths of the poles and, in general, Fermi energy $E_F$, strain-induced splitting $\hbar\Omega_X$ and magnetic splitting $\sim \mu_BB$ all have the same order.

To get the analytic derivations for Fermi polaron energy we consider a large splitting between intra- and intervalley trions, Eq.~\eqref{eq:small}. Thus, we consider two Fermi polaron branches independently. 




The simplified Hamiltonian reads
\begin{equation}
    \hat{\mathcal{H}}^{\text{W}}_{1,2} = \left(-E_{T1,2} + \xi E_F\right)\hat1 + \frac{\Delta E^{(s)}_{\text{FP}}}{2}\hat\sigma_x + \frac{\Delta E^{\text W}_{\text{FP}1,2}}{2}\hat\sigma_z,
\end{equation}
where $\hat1$ is the $2\times 2$ unity matrix and $\hat\sigma_x$, $\hat\sigma_z$ are Pauli matrices, $\Delta E_{\text{FP}}^{(s)}$ is introduced in Eq.~\eqref{eq:DEFPs} and $\Delta E_{\text{FP}1,2}^{\text W}$ is in Eq.~\eqref{eq:dEWFP}. 
The form of the Hamiltonian follows from the symmetry arguments. It combines real magnetic field (term $\propto \sigma_z$) and strain-induced pseudomagnetic field ($\propto \sigma_x$) acting on Fermi polaron pseudospin. The Hamiltonian would be exact for the bare exciton with the different strain- and Zeeman splittings and, in the first order approximation, doping-independent~\cite{glazovExcitonFineStructure2022}.

In this regime effective magnetic field is tilted in $(xz)$ plane, and 
energies of Fermi polaron doublets are
\begin{equation}
    \label{eq:EWs}
    E^\text{W}_{\text s1,2} = -E_{T1,2} + \xi E_F \pm \frac{\sqrt{\left(\Delta E^{\text W}_{\text{FP}1,2}\right)^2 + \left(\Delta E^{(s)}_{\text{FP}}\right)^2}}{2}.
\end{equation}
At $B \to 0$ Fermi polaron states are split by strain into linearly polarized eigenstates. The application of magnetic field mixes these states and leads to the quadratic energy shift (Zeeman effect) of the split states and the redistribution of oscillator strengths. The splitting, therefore, is
\begin{equation}
    \label{eq:DEWs}
    \Delta E^{\text{W}(s)}_{\text{FP}1,2} = \sqrt{\left(\Delta E^{\text W}_{\text{FP}1,2}\right)^2 + \left(\Delta E^{(s)}_{\text{FP}}\right)^2}.
\end{equation}
The exact splitting of the energies corresponding to the poles of Green's function, Eq.~\eqref{eq:Gs}, can be found via numerical solution of equation
\begin{equation}
    \label{eq:EWsExact}
    \left[E - \varepsilon_0^R- \Sigma^R_{\text s,0}(E)\right]\left[E - \varepsilon_0^L- \Sigma^L_{\text s,0}(E)\right]=\Xi^2(E),
\end{equation}
the analogue of equation~\eqref{eq:FPsEnergy} for zero Fermi polaron momentum.

\begin{figure}[ht]
    \centering
    \includegraphics[width=\linewidth]{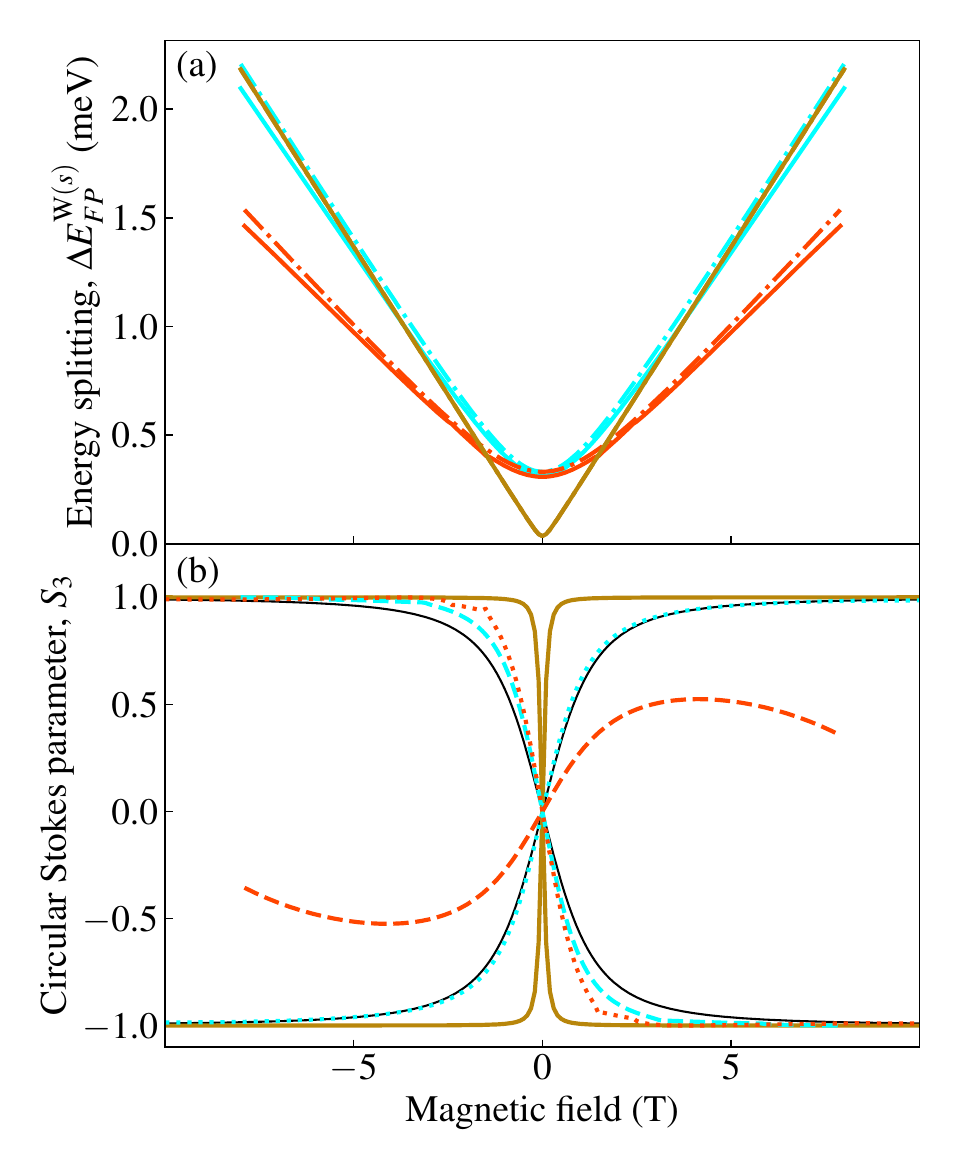}
    \caption{(a) The Fermi polaron radiative doublet splitting and (b) circular polarization in terms of Stokes parameter $S_3$ for strained TMDC monolayers as a function of magnetic field. Cyan lines show intravalley Fermi polaron doublets, while red lines show intervalley Fermi polarons. Solid lines in (a) correspond to the numerical solution of Eq.~\eqref{eq:EWsExact}, while dashed-dotted lines correspond to analytical approximation, Eq.~\eqref{eq:DEWs}. Color lines in (b) correspond to eigenstates of Green's matrix, Eq.~\eqref{eq:Gs}, while thin solid black lines correspond to their approximation with $g = g_X$ for simplicity, Eq.~\eqref{eq:S3W}. Solid brown lines show energy level splitting, Eq.~\eqref{eq:EMos}, and Stokes parameter $S_3$, Eq.~\eqref{eq:S3Mo}, of Fermi polarons in Mo-based TMDCs. Dotted lines correspond to $x$-polarized at $B = 0$ states, dashed lines correspond to $y$-polarized states. $\hbar\Omega_X = 1$~meV, other parameters are the same as in Fig.~\ref{fig:Pol(B)}(c,d) with $N_e = 8\cdot10^{10}$~cm$^{-2}$ and $B_{\text{crit}} = 8$~T.}
    \label{fig:Splitting_hw(B)}
\end{figure}

Figure~\ref{fig:Splitting_hw(B)}(a) shows the comparison between the splitting found numerically by solving Eq.~\eqref{eq:EWsExact} (solid lines) and from simplified two-level model, Eq.~\eqref{eq:DEWs} (dashed-dotted lines). Cyan lines correspond to intravalley Fermi polaron states and red lines~--- to intervalley ones. Both pairs of lines demonstrate good agreement even for the large fields, where valley polarization tends to unity. This can be explained by the dominance of the Zeeman splitting at large magnetic fields, $\Delta E^{\text W}_{\text{FP}1,2} \gg \Delta E^{(s)}_{\text{FP}}$. Brown lines refer to the Mo-based TMDC monolayers discussed below.

The polarization of Fermi polaron states is accordingly determined by the effective magnetic field. The Stokes parameter $S_3$, which describes the circular polarization,
\begin{equation}
    \label{eq:S3W}
    S_3 = \pm\frac{\Delta E^{\text W}_{\text{FP}1,2}}{\sqrt{\left(\Delta E^{\text W}_{\text{FP}1,2}\right)^2 + \left(\Delta E^{(s)}_{\text{FP}}\right)^2}},
\end{equation}
where the `$\pm$' sign corresponds to the two different states within the radiative doublet. The actual state polarization found by the diagonalization of Green's function (Eq.~\eqref{eq:Gs}) at pole energies together with its approximation (Eq.~\eqref{eq:S3W}) are shown in Fig.~\ref{fig:Splitting_hw(B)}(b). The colors are the same as in Fig.~\ref{fig:Splitting_hw(B)}(a), dotted lines correspond to $x$-polarized at $B = 0$ states, dashed lines correspond to $y$-polarized at $B = 0$ states, and solid black lines correspond to simplified approximation~\eqref{eq:S3W} (with $g = g_X$ for simplicity). As the states in the model are fully polarized, at $B = 0$ the states are linearly polarized, while at large magnetic fields they become circularly polarized as in the previous subsection. Note that above the critical magnetic field only two states remain. Three out of four states are in reasonable agreement with the model, while one state remains mostly linearly polarized. The behavior is caused by the complicated interplay of different broadened states and demonstrates the necessity of the complete model.

Above the critical magnetic field electrons in the conduction band remain only in one valley and, thus, only two attractive Fermi polaron states remain. For the sake of certainty, we consider $B > B_{\text{crit}}$, that results in $\tilde{E}_F^l = 0$ and $\tilde{E}_F^r = 2E_F$ with the Fermi polaron energies
\begin{equation}
    E^\text{W}_{\text s,r} = -E_T + 2\xi E_F \pm \frac{\Delta_r}{2},
\end{equation}
where $\Delta_r$ is the renormalized $T_1$~-- $T_2$ splitting introduced in Eq.~\eqref{eq:delta}. In fact, this limit works for arbitrary relation between $|\mu_BB|$, $\hbar\Omega_X$ and $|E_{T1} - E_{T2}|$. The effect of the strain in this case is simply the renormalization of the different Fermi polaron state splitting, $|E_{T1} - E_{T2}|$. Note that for this field range the Zeeman effect is linear.

For Mo-based TMDCs only intervalley trion is bound, thus, the mixing is determined by the mixing with bare exciton state. The self-energy appears to be the same as for the absence of strain, Eq.~\eqref{eq:SigmaMo}, and the mixing terms are only due to the exciton part of wave function
\begin{equation}
    \Xi^\sigma_{\text{Mo}, \bm k}(E_{\bm k}) = \frac{\hbar\Omega_X}{2}.
\end{equation}
We keep the terms $\propto \hbar\Omega_X/E_T$ as they are the dominant strain-induced terms. The exact equations for attractive Fermi polaron energies below and above critical magnetic field are presented in Appendix~\ref{app:energyMo}. It results in a quadratic Zeeman effect at small magnetic fields, $|\mu_BB| \ll E_F$,
\begin{equation}
    \label{eq:EMosApprox}
    E_s^{\text{Mo}} = -E_T + \xi E_F \pm \sqrt{\left(\frac{g_{\text{Mo}}\mu_BB}{2}\right)^2+\left(\alpha E_F\frac{\hbar\Omega_X}{2E_T}\right)^2},
\end{equation}
where $g_{\text{Mo}}$ was derived above in Eq.~\eqref{eq:dEMoFP}. Correspondingly, the Stokes parameter $S_3$ is 
\begin{equation}
    \label{eq:S3Mo}
    S_3 = \pm\frac{g_{\rm Mo}\mu_BB}{\sqrt{\left(g_{\rm Mo}\mu_BB\right)^2 + \left(\alpha E_F \hbar\Omega_X / E_T\right)^2}}.
\end{equation}
Eqs.~\eqref{eq:EMosApprox},~\eqref{eq:S3Mo} have the same form as Eqs.~\eqref{eq:EWs},~\eqref{eq:S3W}, but with the different Fermi polaron $g$ factors and strain-induced splitting.

Brown lines in Fig.~\ref{fig:Splitting_hw(B)} show energy level splitting, Eq.~\eqref{eq:EMos}, and Stokes parameter $S_3$, Eq.~\eqref{eq:S3Mo}, for Fermi polarons in Mo-based TMDCs. The approximate energy splitting, Eq.~\eqref{eq:EMosApprox}, is indistinguishable from the exact expression due to the small strain-induced splitting $\alpha E_F\hbar\Omega_X / E_T = 0.035$~meV for the parameters of Fig.~\ref{fig:Splitting_hw(B)}. Similarly, the crossover magnetic field, at which Zeeman splitting exceeds strain-induced splitting, is sufficiently low, so the states become circularly polarized at magnetic fields of the order of fractions of tesla.

\section{Discussion}
\label{sec:discussion}

\begin{figure*}[ht]
    \centering
    \includegraphics[trim={2cm 0 0 0},clip,width=\linewidth]{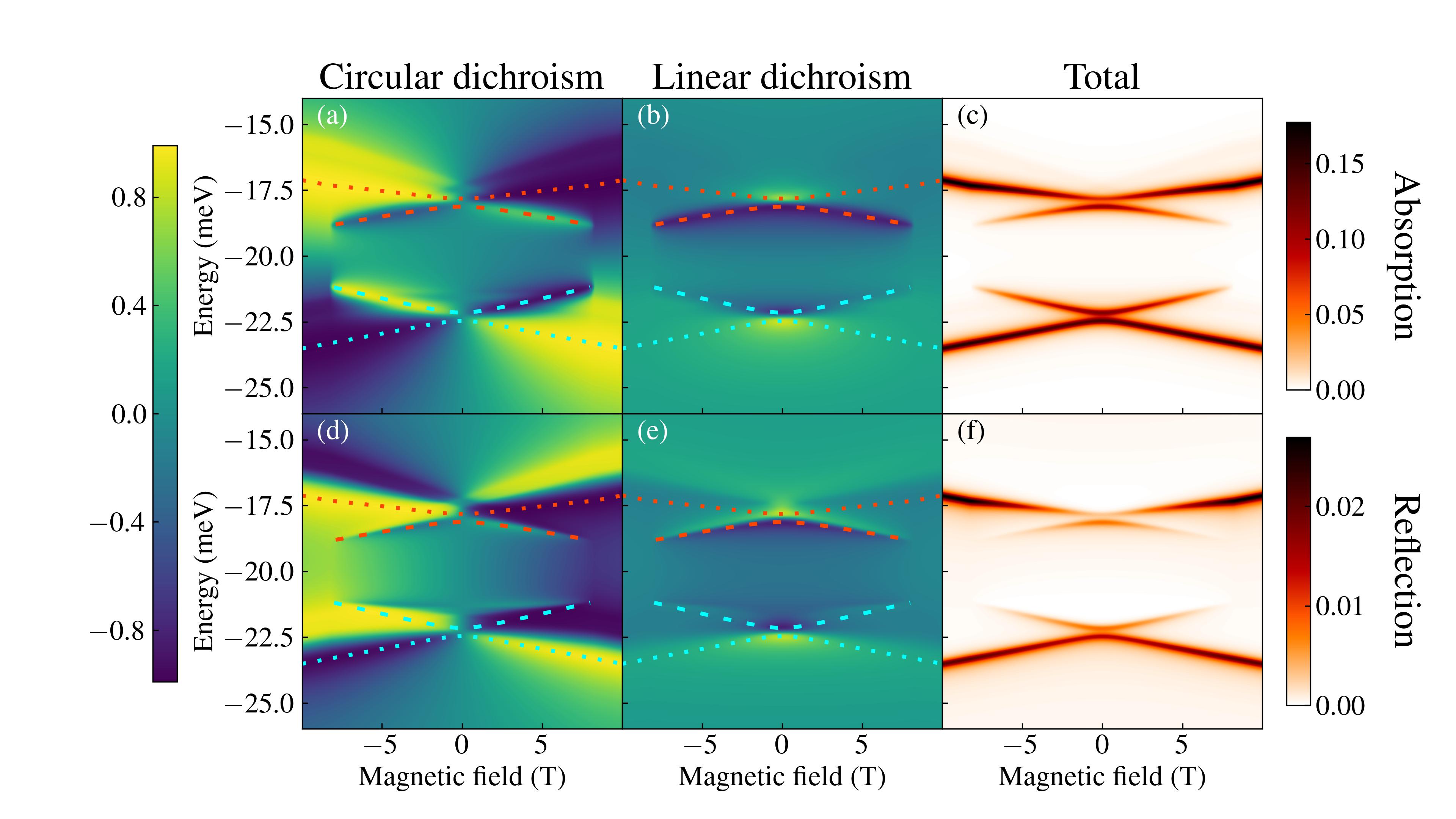}
    \caption{(a)~Circular dichroism of absorption $P_c(E)$, Eq.~\eqref{eq:Pc}, (b)~linear dichroism of absorption $P_l(E) = \left(\mathcal{A}^X-\mathcal{A}^Y\right)/\mathcal{A}$, (c)~absorption spectrum of unpolarized light $\mathcal{A}$, Eq.~\eqref{eq:As}, as a function of magnetic field for strained W-based TMDC. (d)~Circular dichroism of reflection, (e)~linear dichroism of reflection, (f)~reflection spectrum of unpolarized light. Note that the absolute value of absorption is $\sim 10$ times greater than reflection. Color lines describing Fermi polaron energies are the same as in Fig.~\ref{fig:Splitting_hw(B)}. The phenomenological broadening $\hbar\Gamma = 0.1$~meV, the radiative decay rate $\Gamma_0 = 0.65$~ps$^{-1}$~\cite{zhaoStrainControlledAtomicReconstruction2026}, and other parameters are the same as in Fig.~\ref{fig:Splitting_hw(B)} with $N_e = 8\cdot10^{10}$~cm$^{-2}$ and $B_{\text{crit}} = 8$~T.}
    \label{fig:Pol_hw(B)}
\end{figure*}

In the general case, when both Zeeman and pseudo-Zeeman splittings are present, the Fermi polaron eigenstates are elliptically polarized (see Fig.~\ref{fig:Splitting_hw(B)}(b)), and the optical response exhibits both circular and linear dichroism in reflection, transmission and absorption. Figure~\ref{fig:Pol_hw(B)} demonstrates the optical properties of a W-based TMDC monolayer for the same parameters as in Figs.~\ref{fig:Pol(B)}(c,d), and~\ref{fig:Splitting_hw(B)}, and experimental value of radiative decay rate $\Gamma_0 = 0.65$~ps$^{-1}$~\cite{zhaoStrainControlledAtomicReconstruction2026}. Panels (a)~-- (c) show the absorption spectra while panels (d)~-- (f) show the reflection spectra. Panels (a,d) demonstrate circular dichroism, panels (b,e) show linear dichroism and panels (c,f) present absorption and reflection spectra of unpolarized light, respectively. In this set of parameters the absorption is one order of magnitude greater than reflection. Color lines in dichroism panels show eigenenergies of Fermi polaron states with the same colorcode as in previous figures, dotted lines denote $x$-polarized at $B = 0$ states and dashed lines denote $y$-polarized at $B = 0$ ones.

The nonzero strain-induced splitting at $B = 0$ together with quadratic Zeeman energy shift are clearly visible at small magnetic field $\lesssim 1$~T. At such fields Fermi polaron states are linearly polarized and there is no circular dichroism, while circularly polarized light exhibits polarization conversion in reflection and transmission due to the strain-induced anisotropy of the system. At elevated magnetic field its effects become dominant and the states become mostly circularly polarized (except for the $y$-polarized intervalley Fermi polaron state due to the interplay of the nearby states). Circular dichroism plots together with absorption plots demonstrate clear similarity with Fig.~\ref{fig:Pol(B)}(c,d) with the redistribution of oscillator strengths seen in absorption and reflection spectra. 
In addition to the linear polarization conversion TMDC, monolayer should demonstrate the polarization in luminescence due to the thermal population of different state in radiative doublet even under unpolarized light excitation. This effect increases with the increase of the magnetic field due to the increase in the value of splitting. At large magnetic field above the critical value $B > B_{\text{crit}} = 8$~T electrons in conduction band become fully valley-polarized and again only two states remain.

In Mo-based TMDC monolayers the situation, in general, is similar to that in W-based systems. The main difference is the smaller strain-induced splitting due to the admixing of split bare exciton state. It results in much lower crossover magnetic field between linearly and circularly polarized states. However, the small splitting complicates the possible experimental observation of the interplay between Zeeman and pseudo-Zeeman splittings due to the finite broadening of Fermi polaron lines. 

We note that Zeeman and pseudo-Zeeman effects and the optical properties of p-doped TMDC monolayers are similar to the n-doped Mo-based monolayer independently of the transition metal. Spin-orbit splitting in the valence band of TMDC monolayers is comparatively big, so the resident holes occupy only top valence subband relevant for the bright exciton. Therefore, due to the exchange interaction only intervalley attractive Fermi polaron is bound resulting in similar to the Mo-based TMDC monolayers behavior.

\section{Conclusion}
\label{sec:conclusion}

We develop a theory of attractive Fermi polarons in TMDC monolayers in an external magnetic field, considering both isotropic monolayers and monolayers under uniaxial strain. Unlike trion model, Fermi polaron model takes into account correlations with the resident charge carriers in the Fermi sea. The valley-dependent Zeeman shifts of the conduction band minima induce valley polarization of TMDC monolayers proportional to magnetic field strength. We demonstrate that resident charge carrier density affects the attractive Fermi polaron binding energy emerging the density-independent at small magnetic fields renormalization of $g$~factors with the different signs for intravalley and intervalley attractive Fermi polarons. The Zeeman effect splits attractive Fermi polaron states into circularly polarized eigenstates. Above the critical magnetic field the resident charge carriers become fully valley polarized and only one Fermi polaron state in the radiative doublet remains. At the critical field separating partially and fully valley-polarized regimes, the density dependence of the Zeeman splitting exhibits a kink.

When an external magnetic field and uniaxial strain are applied simultaneously, the Zeeman and pseudo-Zeeman effects compete and hybridize the Fermi polaron states. We propose a theory for attractive Fermi polaron states in the presence of external magnetic field and uniaxial strain, that takes into account both field-induced valley polarization and strain-induced valley mixing effects. The derived tensor Green's function of Fermi polaron radiative quadruplet allows us to find Fermi polaron eigenenergies and eigenstates Stokes parameters. The simplified model taking into account Zeeman and pseudo-Zeeman splittings provides qualitative results in agreement with numerics. At small magnetic fields the Fermi polaron state polarization is close to linear strain-induced and the Zeeman shift is quadratic. At elevated magnetic fields the Zeeman effect dominates over the strain-induced splitting resulting in linear in magnetic field Zeeman splitting with eigenstates evolving from elliptically polarized to nearly circuarly polarized. Above the critical magnetic field electrons become fully valley polarized and the Fermi plaron states are fully circularly polarized. We calculate the absorption and reflection spectra and linear and circular dichroism in the energy range of attractive Fermi polarons, demonstrating the ability to control the energy and polarization of eigenstates and their optical response.

The developed theory proposes ways for experimental control of attractive Fermi polarons with continuously tunable energy splitting and the polarization of the states within each radiative doublet in experiments similar to those reported in~\cite{jasinskiStrainInducedLifting2022,yagodkinFermiPolaronsStraininduced2025}. Similar effects can be observed in van der Waals heterostructures and conventional semiconductor quantum wells. Possible directions for future research include valley coherence and entanglement of Fermi-polaron states, as well as their transport properties under circularly and linearly polarized optical excitation.

\section*{Acknowledgments}
We thank M.M. Glazov for helpful discussions. The author gratefully acknowledge RSF Project No. 23-12-00142$\Pi$ and the BASIS foundation for financial support.




\appendix

\section{Zeeman effect self-energies}
\subsection{W-based TMDC}
\label{app:SigmaW}

To calculate the self-energy, we estimate
\begin{equation}
    S^{\sigma\tau}_0(\bm q) = 
        -\mathcal{D}\ln\dfrac{E_X}{-\left(E^\sigma_{0} - \eps^\sigma_0\right) - \frac{M_X\hbar^2q^2}{2M_eM_T} + \frac{M_T}{M_X}\tilde{E}_F^\tau}.
\end{equation}

Finally, using the energy scale assumptions~\eqref{eq:energy_scales}, we can estimate the self-energy
\begin{multline}
    \label{eq:SigmaAa:int}
    \Sigma^{\sigma\tau}_0\left(E^\sigma_0\right) = \left(\frac{M_T}{M_X}\right)^2E_{T,\sigma\tau} \\ \times \ln\left(1 + \frac{M_X}{M_T}\frac{\tilde{E}_F^\tau}{E^\sigma_0 - \eps^\sigma_0 + E_{T,\sigma\tau} - \frac{M_T}{M_X}\tilde{E}^\tau_F}\right).
\end{multline}

\subsection{Mo-based TMDC}
\label{app:SigmaMo}

The exciton self-energy in Mo-based TMDCs in $\bm K_+$-valley is
\begin{subequations}
    \label{eq:SigmaMo}
    \begin{multline}
        \Sigma^{R}_{\text{Mo}, 0}\left(E^R_{0}\right) = \sum_{\bm q}\frac{V_{2}}{1 - V_{2}S^{Rl}_{0}(\bm q)} = \left(\frac{M_T}{M_X}\right)^2E_{T} \\ \times \ln\left(1 + \frac{M_X}{M_T}\frac{\tilde{E}_F^l}{E^R_0 - \eps^R_0 + E_{T} - \frac{M_T}{M_X}\tilde{E}^l_F}\right),
    \end{multline}
    where we omit index $2$ for the single existing intervalley trion energy $E_T$ and write the expressions at zero wavevector. For exciton in $\bm K_-$ valley the self-energy is
    \begin{multline}
        \Sigma^{L}_{\text{Mo}, 0}\left(E^L_{0}\right) = \left(\frac{M_T}{M_X}\right)^2E_{T} \\ \times \ln\left(1 + \frac{M_X}{M_T}\frac{\tilde{E}_F^r}{E^L_0 - \eps^L_0 + E_{T} - \frac{M_T}{M_X}\tilde{E}^r_F}\right).
    \end{multline}
\end{subequations}

\section{Strained TMDC monolayers}
\subsection{Self-energies in W-based TMDC}
\label{app:SigmaSW}
Using the energy scales from Eq.~\eqref{eq:energy_scales}, we derive the exciton self-energy taking into account both magnetic field and strain-induced valley-mixing effects
\begin{multline}
    \Sigma^{\sigma\tau}_{\text s,0}\left(E\right) = \left(\frac{M_T}{M_X}\right)^2\frac{E_{T,\sigma\tau}}{2}\\\times\left[\left(1 + \frac{\delta_{\sigma'\tau}}{\Delta_\tau}\right)\ln\left(1 + \frac{\frac{M_X}{M_T}\tilde{E}_F^\tau}{E + E_T - \frac{\Delta_\tau}{2} - \frac{M_T}{M_X}\tilde{E}_F^\tau}\right)\right. \\ + \left.\left(1 + \frac{\delta_{\sigma\tau}}
    {\Delta_\tau}\right)\ln\left(1 + \frac{\frac{M_X}{M_T}\tilde{E}_F^\tau}{E + E_T + \frac{\Delta_\tau}{2} - \frac{M_T}{M_X}\tilde{E}_F^\tau}\right)\right],
    \label{eq:SigmasAa:int}
\end{multline}
where we introduce the renormalized difference between trion energies
\begin{equation}
    \label{eq:delta}
    \delta_{\sigma\tau} = E_{T,\sigma\tau} - E_{T,\sigma'\tau} - \eps_0^\sigma + \eps_0^{\sigma'}, \quad \Delta_\tau = \sqrt{\delta_{\sigma\tau}^2 + \left(\hbar\Omega_X\right)^2}.
\end{equation}
Note that the sign of energy difference is determined by the exciton valley $\sigma$, $\delta_{\sigma\tau} = -\delta_{\sigma'\tau}$, thus, $\Delta_\tau$ depends only on electron valley and is independent on the exciton valley. At negligible strain, $\hbar\Omega_X \ll |\delta_{\sigma\tau}|$, we get $|\delta_{\sigma\tau}| \to \Delta_\tau$ and only one summand in Eq.~\eqref{eq:SigmasAa:int} remains, that correlates with previous derivations, Eq.~\eqref{eq:SigmaAa:int}.

In the dominant in trion binding energy~$E_T$ term, the mixing-induced terms depend only on electron valley index~$\tau$ [$\Xi^{\sigma\tau}_{0}(E) = \Xi^{\sigma'\tau}_0(E)\equiv \Xi^\tau_0(E)$]
\begin{multline}
    \Xi^\tau_{0}(E) = \left(\frac{M_T}{M_X}\right)^2E_T\frac{\hbar\Omega_X}{2\Delta_\tau}\\\times\left[\ln\left(1 + \frac{M_X}{M_T}\frac{\tilde{E}_F^\tau}{E + E_T - \frac{\Delta_\tau}{2} - \frac{M_T}{M_X}\tilde{E}_F^\tau}\right)\right. \\ - \left.\ln\left(1 + \frac{M_X}{M_T}\frac{\tilde{E}_F^\tau}{E + E_T + \frac{\Delta_\tau}{2} - \frac{M_T}{M_X}\tilde{E}_F^\tau}\right)\right],
\end{multline}
and in the same approximation the total mixing terms are equal
\begin{equation}
    \Xi^R_0(E) = \Xi^L_0(E) \equiv \Xi(E) = \frac{\hbar\Omega_X}{2} + \Xi^\tau_0(E) + \Xi^{\tau'}_0(E).
\end{equation}
This symmetry relation leads to the symmetry of Green's function, Eq.~\eqref{eq:Gs}, $\mathcal G^{RL}_s(E) = \mathcal G^{LR}_s(E)$.

\subsection{Fermi polaron energies in Mo-based TMDC}
\label{app:energyMo}
The attractive Fermi polaron doublet energies for Mo-based TMDCs in a magnetic field below the critical value, $|B| < B_{\text{crit}}$, are
\begin{multline}
    \label{eq:EMos}
    E^{\text{Mo}}_{\text s} = -E_T + \xi E_F \pm \left[\left(\eps_0^L+\xi\frac{\tilde{E}_F^r - \tilde{E}_F^l}{2}\right)^2\right. \\ + \alpha^2\tilde{E}_F^r\tilde{E}_F^l\left(\frac{\hbar\Omega_X}{2E_T}\right)^2 \\ \left. -\alpha\left(\eps_0^L+\xi\frac{\tilde{E}_F^r - \tilde{E}_F^l}{2}\right)\left(\tilde{E}_F^r - \tilde{E}_F^l\right)\left(\frac{\hbar\Omega_X}{2E_T}\right)^2\right]^{1/2}.
\end{multline}
Above the critical magnetic field, $|B| > B_{\text{crit}}$, at fully valley-polarized Mo-based TMDC only one Fermi polaron state remains with the energy
\begin{multline}
    E_s^{\text{Mo}} = -E_T + \xi E_F - \left[\left(\frac{g_X\mu_B|B|}{2} + \xi E_F\right)\right. \\ \left.\times\left(\frac{g_X\mu_B|B|}{2} + \xi E_F - 2\alpha E_F\left(\frac{\hbar\Omega_X}{2E_T}\right)^2\right)\right]^{1/2},
\end{multline}
coinciding with Eq.~\eqref{eq:EMo} due to the parametrically small factor in the last term.

\bibliography{biblio}
\end{document}